\documentclass[runningheads]{llncs}
\usepackage[T1]{fontenc}
\usepackage{url}
\usepackage[table]{xcolor}
\usepackage{pgfplots}
\pgfplotsset{compat=1.18}

\usepackage{graphicx}
\usetikzlibrary{patterns}
\usepackage{placeins}

\begin{document}
	\title{Generative Artificial Intelligence and the Knowledge Gap Perspective:}
	\subtitle{Toward a New Form of Informational Inequality}
	\author{Raphael Morisco}
	\authorrunning{Morisco}
	\institute{Independent Scholar \\
		Media, Technology, and Security Research \\
		Karlsruhe, Germany\\
		Formerly with Karlsruhe Institute of Technology (KIT)
	}
	\maketitle
	\begin{abstract}
		The knowledge gap hypothesis suggests that the diffusion of information tends to increase rather than reduce social inequalities. Subsequent research on the digital divide has extended this perspective by focusing on unequal access to and use of digital technologies. The emergence of generative artificial intelligence raises the question of whether these frameworks remain sufficient to describe current forms of informational inequality. While access to AI systems is increasingly widespread, differences may arise in how users engage with AI-generated content. This paper proposes a theoretical extension of the knowledge gap perspective by arguing that generative AI shifts the focus from access and usage to the critical evaluation of information. It is assumed that individuals with higher levels of education are more likely to question and contextualize AI-generated outputs, whereas individuals with lower levels of education may rely more directly on them. The contribution is conceptual and does not present empirical findings. It aims to provide a framework for future research on the relationship between education, AI use, and knowledge inequality.
		
	
	\end{abstract}
	
	\section{Introduction}
	
	"With the increasing diffusion of information, segments of the population with higher socioeconomic status tend to acquire this information at a faster rate than lower-status segments" \cite{tichenor1970}, see also \cite[p.~12]{zillien2014}. This assumption, commonly referred to as the knowledge gap hypothesis, has shaped research on informational inequality for several decades \cite{bonfadelli2017}. Subsequent developments in the field, particularly research on the digital divide, have extended this perspective by focusing on disparities in access to and use of digital technologies \cite{mauch2016,zillien2014}. In recent years, the emergence of generative artificial intelligence has introduced new forms of interaction with information. Systems that generate text, images, or other forms of content are increasingly accessible to a broad range of users. As a result, the question arises whether established frameworks such as the knowledge gap perspective and the digital divide remain sufficient to describe current forms of informational inequality. While earlier approaches primarily addressed differences in access to information or in patterns of technology use, generative AI systems are characterized by their capacity to produce information. This shifts the point of interaction from retrieval to interpretation. Users are no longer only required to locate and access information, but also to assess, contextualize, and evaluate outputs that are generated by technical systems.
	
	The present paper takes this development as a starting point and proposes a theoretical extension of the knowledge gap perspective. It is argued that generative AI introduces a further dimension of informational inequality that is not adequately captured by existing models. In particular, differences may arise in how users engage with AI-generated content. It is assumed that individuals with higher levels of education are more likely to question and contextualize such outputs, whereas individuals with lower levels of education may rely more directly on them. The paper is structured as follows. Section~2 outlines the theoretical background by revisiting the knowledge gap hypothesis and subsequent research on the digital divide, followed by a discussion of more recent approaches that point toward a modified understanding of informational inequality. Section~3 examines the specific characteristics of generative artificial intelligence and its implications for existing frameworks. Based on this, Section~4 develops a conceptual extension of the knowledge gap perspective. Section~5 discusses implications for education, society, and future research. Finally, Section~6 addresses limitations and outlines directions for further empirical investigation.

	\section{Theoretical Background}
	
	\subsection{The Knowledge Gap Perspective}
	
	The knowledge gap hypothesis, originally formulated by Tichenor, Donohue, and Olien, describes the tendency for information diffusion to increase rather than reduce social inequalities. It assumes that individuals with higher socioeconomic status are able to acquire information at a faster rate than those with lower status \cite{tichenor1970}, see also \cite[p.~12]{zillien2014}. At its core, the knowledge gap perspective is based on the assumption of an excluding effect of media-based information dissemination, which is closely linked to the socioeconomic status of individuals within a society \cite[pp.~12]{zillien2014}. Over the past decades, the field has developed into a differentiated body of research comprising various analytical approaches and empirical designs \cite[pp.~241--244]{bonfadelli2017}. At the same time, the theoretical foundations of the knowledge gap hypothesis have been subject to critique. In particular, the concept of knowledge itself has been described as insufficiently specified, and the methodological approaches used in empirical studies have been considered in need of further refinement \cite[pp.~93--100]{zillien2014}. Despite these limitations, the knowledge gap hypothesis remains an established framework for analyzing informational inequality.
	
	\subsection{The Digital Divide}
	
	As a continuation of knowledge gap research, the concept of the digital divide shifts the focus from differences in knowledge acquisition to disparities in access to and use of digital information and communication technologies \cite[pp.~74--77]{zillien2014}. While the knowledge gap perspective emphasizes the role of socioeconomic status in shaping information acquisition, digital divide research is primarily concerned with unequal opportunities to access and utilize digital technologies \cite[p.~211]{mauch2016}. Within this field, various models have been developed to capture different dimensions of digital inequality, including access, usage, and competencies. These dimensions are often operationalized through indicators such as frequency of internet use, technical skills, and the types of content accessed online \cite[p.~81]{zillien2014}. However, similar to knowledge gap research, the study of the digital divide has also been subject to methodological critique. In particular, the lack of standardized operationalizations and measurement approaches has made it difficult to compare findings across studies \cite[pp.~275--276]{marr2010}. This has led to calls for more integrative and multidimensional approaches to the analysis of digital inequality \cite[p.~217]{mauch2016}.
	
	\subsection{Toward a New Digital Divide}
	
	More recent approaches suggest that neither socioeconomic status nor access to technology alone sufficiently explain contemporary forms of digital inequality. This becomes particularly evident in the context of higher education, where access to digital infrastructure is largely standardized and does not constitute a primary source of differentiation. In this context, it has been argued that a new form of digital divide emerges from differences in the structure of knowledge acquisition itself. Rather than being determined by access, inequalities arise from the distribution of domain-specific knowledge across different fields of study. For instance, knowledge related to information technology and IT security is often concentrated in specific disciplines, while remaining largely absent from others \cite[pp.~632, 636]{morisco2019}. Accordingly, the focus shifts from access and basic usage to more differentiated forms of knowledge and competencies. Digital inequality is thus increasingly shaped by educational structures and disciplinary boundaries, which determine which forms of knowledge are made available to different groups of individuals. This development points toward a modified understanding of informational inequality, in which differences persist despite comparable levels of access to digital technologies. It is this shift that provides the theoretical basis for examining further transformations in the context of generative artificial intelligence.

	\section{Generative Artificial Intelligence and Informational Inequality}
	
	The development of generative artificial intelligence introduces a new form of interaction with information that differs in several respects from earlier digital technologies. While traditional information systems primarily support the retrieval and distribution of existing content, generative AI systems are capable of producing new content in response to user input. This includes text, images, and other forms of representation that are increasingly integrated into everyday information practices. As a result, the role of the user changes. Instead of searching for and selecting information from existing sources, users interact with systems that generate responses dynamically. This alters the conditions under which information is accessed and processed. The interaction is no longer limited to locating relevant information, but extends to interpreting and assessing outputs that are produced by technical systems.
	
	From the perspective of informational inequality, this shift raises the question of whether existing frameworks remain sufficient. The knowledge gap perspective and research on the digital divide have primarily focused on differences in access to information and in patterns of technology use. However, these approaches do not fully account for situations in which information is not merely accessed, but generated. At the same time, access to generative AI systems is becoming increasingly widespread. Many applications are publicly available and can be used without specialized technical knowledge. This suggests that access, as a central explanatory factor in digital divide research, may be of decreasing relevance in this context. Instead, differences may arise in how users engage with AI-generated content. In particular, the use of generative AI requires the ability to interpret outputs, to assess their plausibility, and to relate them to existing knowledge. Since AI-generated content may contain inaccuracies, omissions, or context-dependent limitations, users are required to exercise a certain degree of critical judgment. This introduces an additional layer of complexity that is not fully captured by existing models of informational inequality. Taken together, these developments indicate that generative artificial intelligence may transform the conditions under which knowledge is acquired and processed. This provides the basis for reconsidering established perspectives on informational inequality and for developing a conceptual extension that accounts for the specific characteristics of AI-mediated information.

	\section{A Theoretical Extension: The Generative AI Knowledge Gap}
	
	The preceding discussion has shown that generative artificial intelligence alters the conditions under which information is accessed and processed. While existing frameworks such as the knowledge gap perspective and the digital divide account for differences in access and usage, they do not fully capture situations in which information is generated dynamically and must be interpreted by users. Against this background, it is possible to conceptualize a further development of informational inequality that emerges in the context of generative AI. This paper proposes a theoretical extension of the knowledge gap perspective, in which differences arise not primarily from access to information or from patterns of technology use, but from the way in which users engage with AI-generated content.

	\subsection{From Access to Evaluation}
	
	In earlier stages of research, informational inequality has been closely linked to disparities in access to information and communication technologies. With the increasing diffusion of digital infrastructures, this focus has shifted toward differences in usage and competencies. The development of generative AI suggests a further shift, in which the central point of differentiation lies in the evaluation of information. Since AI systems generate outputs that are not directly tied to identifiable sources, users are required to assess the plausibility and relevance of the information provided. This introduces a form of informational interaction in which evaluation becomes a central component. Differences in the ability to perform such evaluations may therefore constitute a new dimension of inequality.

	\subsection{Critical Engagement with AI-Generated Content}
	
	The use of generative AI involves varying forms of engagement with system outputs. In this context, a distinction can be made between more critical and less critical modes of use. Critical engagement includes practices such as questioning the correctness of outputs, comparing results with alternative sources, and reflecting on the limitations of the system. Less critical forms of use, by contrast, may involve a more direct acceptance of generated content without further verification. It can be assumed that these different modes of engagement are not evenly distributed across users. In particular, educational background may play a role in shaping how individuals interact with AI-generated information. Individuals with higher levels of education may be more likely to adopt a critical stance, whereas individuals with lower levels of education may rely more directly on the outputs provided.
	
	\subsection{Epistemic Dimensions of AI Use}
	
	The differences outlined above point toward a broader set of competencies that can be described as epistemic in nature. These include the ability to interpret information in context, to recognize potential limitations or inaccuracies, and to relate new information to existing knowledge structures. In contrast to domain-specific forms of knowledge, these competencies are not tied to a particular field, but are relevant across different contexts of AI use. As generative AI systems become more widely integrated into everyday information practices, such competencies may become increasingly important in shaping how knowledge is acquired and processed. Taken together, these considerations suggest that generative AI introduces a new dimension of informational inequality. Building on the knowledge gap perspective, this can be understood as a shift from differences in access and usage toward differences in the evaluation and interpretation of information. This provides the basis for conceptualizing what may be described as a generative AI knowledge gap.	
	
	\section{Implications}
	
	The proposed extension of the knowledge gap perspective has implications for different areas, including education, society, and future research. By shifting the focus from access and usage to the evaluation of information, generative artificial intelligence introduces new challenges that are not fully addressed by existing approaches to informational inequality.
	
	\subsection{Implications for Education}
	
	In the context of education, the increasing relevance of generative AI highlights the importance of competencies that go beyond basic digital skills. While access to technology and the ability to use digital tools remain important, the ability to critically assess AI-generated content becomes a central requirement. This suggests that educational approaches may need to place greater emphasis on the development of critical and reflective skills. In particular, the ability to question information, to evaluate its plausibility, and to relate it to existing knowledge may become increasingly relevant. Such competencies are not limited to specific disciplines, but apply across different fields of study.
	
	\subsection{Societal Implications}
	
	At a broader societal level, the integration of generative AI into everyday information practices may contribute to new forms of inequality. If differences in the evaluation of AI-generated content are unevenly distributed, this may affect how individuals form opinions and make decisions. In this context, the increasing availability of AI-generated information may also influence the perceived reliability of information. If users rely on generated outputs without critical assessment, this may lead to the reinforcement of existing misconceptions or to the uncritical acceptance of inaccurate information. Conversely, more critical forms of engagement may enable users to better navigate complex information environments.
	
	\subsection{Implications for Research}
	
	The considerations outlined in this paper point toward several directions for future research. In particular, it may be necessary to further investigate how different groups of users interact with generative AI systems and how these interactions relate to existing forms of informational inequality. This includes examining the role of education in shaping patterns of AI use, as well as identifying relevant competencies that influence how users interpret and evaluate generated content. In addition, future research may explore how these dynamics evolve over time as generative AI systems become more widely adopted and integrated into various domains.	
	
	\section{Limitations and Future Research}
	
	The present contribution is subject to several limitations. First, the paper is conceptual in nature and does not provide empirical evidence for the proposed extension of the knowledge gap perspective. The arguments are based on theoretical considerations and on the interpretation of existing research, rather than on systematic empirical analysis. Second, the focus on education as a potential explanatory factor simplifies the complexity of processes that shape how individuals interact with generative AI systems. Other variables, such as prior knowledge, digital competencies, or contextual factors, may also influence patterns of use and evaluation, but are not examined in detail here. Third, the analysis is limited to a general consideration of generative AI and does not differentiate between specific systems, application contexts, or user groups. Given the diversity of current AI applications, it is likely that patterns of interaction vary across different domains. Against this background, further research is required to examine the assumptions outlined in this paper. In particular, empirical studies may investigate how users with different educational backgrounds engage with AI-generated content, how they assess its reliability, and which competencies are relevant for critical interaction. Experimental and survey-based approaches may be suitable to capture such differences in a systematic manner. In addition, longitudinal perspectives may be useful to analyze how patterns of AI use develop over time and how they relate to broader transformations in informational practices. Such studies may contribute to a more comprehensive understanding of how generative artificial intelligence reshapes existing forms of informational inequality.

	\section{Conclusion}
	
	This paper has examined the implications of generative artificial intelligence for established perspectives on informational inequality. Building on the knowledge gap hypothesis and subsequent research on the digital divide, it has been argued that generative AI introduces a further shift in the conditions under which information is accessed and processed. While earlier approaches focused on differences in access to information and in patterns of technology use, the increasing relevance of AI-generated content highlights the importance of evaluation and interpretation. In this context, differences in how users engage with such content may constitute a new dimension of inequality. By proposing a conceptual extension of the knowledge gap perspective, this paper has outlined a framework for understanding these developments. Although the argument remains theoretical, it points toward the need to reconsider existing models of informational inequality in light of ongoing technological change. Further empirical research is required to examine the extent to which these considerations can be observed in practice and how they evolve over time.

\end{document}